\documentclass[graybox]{svmult}

\usepackage{type1cm}        
\usepackage{makeidx}         
\usepackage{graphicx}        
\usepackage{multicol}        
\usepackage[bottom]{footmisc}

\usepackage{newtxtext}       %
\usepackage[varvw]{newtxmath}       

\makeindex             

\begin{document}

\title*{Intelligent matter consisting of active particles}
\author{Julian Jeggle and Raphael Wittkowski}
\institute{Julian Jeggle 
\newline 
Institute of Theoretical Physics, Center for Soft Nanoscience, University of Münster, 48149 Münster, Germany
\vskip1ex 
Raphael Wittkowski 
\newline
Department of Physics, RWTH Aachen University, 52074 Aachen, Germany
\newline 
DWI -- Leibniz Institute for Interactive Materials, 52074 Aachen, Germany
\newline 
Institute of Theoretical Physics, Center for Soft Nanoscience, University of Münster, 48149 Münster, Germany}
%
%
\maketitle

\abstract*{In this book chapter, we review how systems of simple motile agents can be used as a pathway to intelligent systems. It is a well known result from nature that large groups of entities following simple rules, such as swarms of animals, can give rise to much more complex collective behavior in a display of \textit{emergence}. This begs the question whether we can emulate this behavior in synthetic matter and drive it to a point where the collective behavior reaches the complexity level of intelligent systems. Here, we will use a formalized notion of ``intelligent matter'' and compare it to recent results in the field of active matter. First, we will explore the approach of \textit{emergent computing} in which specialized active matter systems are designed to directly solve a given task through emergent behavior. This we will then contrast with the approach of \textit{physical reservoir computing} powered by the dynamics of active particle systems. In this context, we will also describe a novel reservoir computing scheme for active particles driven ultrasonically or via light refraction.}

\abstract{In this book chapter, we review how systems of simple motile agents can be used as a pathway to intelligent systems. It is a well known result from nature that large groups of entities following simple rules, such as swarms of animals, can give rise to much more complex collective behavior in a display of \textit{emergence}. This begs the question whether we can emulate this behavior in synthetic matter and drive it to a point where the collective behavior reaches the complexity level of intelligent systems. Here, we will use a formalized notion of ``intelligent matter'' and compare it to recent results in the field of active matter. First, we will explore the approach of \textit{emergent computing} in which specialized active matter systems are designed to directly solve a given task through emergent behavior. This we will then contrast with the approach of \textit{physical reservoir computing} powered by the dynamics of active particle systems. In this context, we will also describe a novel reservoir computing scheme for active particles driven ultrasonically or via light refraction.}

\section{Introduction}
\label{sec:1}
The term ``intelligence'' is typically attributed to biological entities as an extreme form of \textit{emergent} behavior as the (relatively) simple biochemical and biophysical interactions between neurons give rise to dramatically more complex psychological phenomena. Many approaches of artificial intelligence thus try to create synthetic analogs of these biological structures. An example are artificial neural networks which derive from a strongly simplified view on neuron activity \cite{Schmidhuber2015}. However, we can also find a form of intelligence in other systems with emergent behavior such as swarms of bacteria \cite{BenJacob2009}, schools of fish \cite{Pitcher1998}, or even groups of humans \cite{Kerr2004}. These systems exhibit collective behavior that transcends the capabilities of each individual and allows the system as a whole to adapt to its environment in an intelligent manner (e.g., to avoid a perceived threat). The observation of this ``swarm intelligence'' in natural systems is an inspiration for an alternative path to artificial intelligence based on systems of synthetic motile agents or, as it is more commonly known as, synthetic \textit{active matter} \cite{Ramaswamy2010}.

As an inherently nonequilibrium system, active matter exhibits a wide range of collective effects not found in nonactive (i.e., passive) materials. Examples include motility-induced phase separation \cite{Cates2015,Jeggle2019}, swarming \cite{Wensink2012}, or self-assembly into complex structures \cite{Mallory2018}. It is the persistent energy flux caused by the energy dissipation of the active agents that permits these advanced behavior patterns in the first place. This correlates well with the aim to produce intelligent materials as such an energy flux is indeed necessary for adaptive (and ultimately intelligent) materials \cite{Walther2019} and has inspired much work into bringing about artificial intelligence on the platform of active matter.

In this chapter, we will give an overview of recent advances in this topic. However, given the large size of the active matter community we will limit ourselves to active matter systems on the microscale where the individual agents do not possess any intelligence themselves nor are they controlled by an external (artificial) intelligence. In other words, we restrict ourselves to systems where intelligence emerges as a purely collective phenomenon. This is justified as current manufacturing capabilities severely limit the complexity of micro- and nanoparticles while systems with larger agents (e.g., robotic swarms) suffer from obvious scaling problems. It must be stressed that this is not a principal limitation, however, as is clearly demonstrated by the remarkable complexity of microorganisms found in nature. Readers interested in the \textit{application} of machine learning on active matter systems are best referred to the chapter of Löwen and Liebchen as well as the chapter by Volpe in this book.

\begin{figure}
    \centering
    \includegraphics{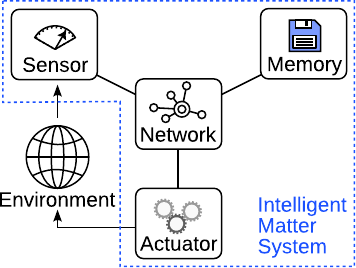}
    \caption{Overview of the four components of intelligent matter as defined by Ref. \cite{Kaspar2021} and their interaction with the environment of the system. The dashed blue outline indicates the boundary of the material system. The black lines denote signal pathways along which information is passed along. They can allow either directional, one-way information flow (direction indicated by an arrow head) or bidirectional, two-way information flow (no arrow heads). The sensor component serves as an information input to the system while the actuating component allows information output. The network component allows for information exchange within the material itself between the other three components. It should be noted that the information pathways inside the material are not directional as internal feedback between components is possible.}
    \label{fig:fig1}
\end{figure}

Before diving into the concrete approaches of this field, it is well advised to first establish a common understanding of what the defining properties of an intelligent system are. In this chapter, we will use the definition proposed by Kaspar et al.\ in Ref. \cite{Kaspar2021}. There, it is argued that four key functionalities are  required\footnote{It should be noted that the presence of these components does not immediately prove that the system is intelligent. Such a proof typically requires the system to actually solve a complex task that requires adaptivity and learning.} for a class of materials to be capable of ``intelligence'', which is here understood as the ability of a system to adapt to and learn from its environment. It should be noted that this definition is not limited to material systems comprised of individual agents as is the case for active particle systems, but can also be applied to materials with tighter integration such as molecular systems (e.g., reaction networks \cite{Goudarzi2013}). In multiagent systems, the functional requirements described in the following only need to be realized by the material as a whole and not necessarily by each individual agent.

First, there needs to be a sensing component that allows the system to both receive information about its environment and harvest energy for its internal workings. Secondly, there needs to be an actuation component that changes the material properties\footnote{Typically, a change in the behavior on the microscopic level will also lead to a different physical behavior of the system as a whole.} and thus effectively providing information output to the environment. The third required component must be capable of information storage, i.e., memory, thus allowing the system to keep track of previous information and enabling learning. Finally, the fourth component is the information network that connects the previous three components. For learning, it is necessary that this information network also permits feedback, i.e., the actuation result of the system must also be available as an information input. A schematic overview of these components is shown in Fig. \ref{fig:fig1}. Besides \textit{intelligent} matter, Kaspar et al.\ also formalize the terms \textit{responsive} matter as matter that has only the sensing and actuation components as well as \textit{adaptive} matter as a matter system that has sensing, actuation, and network components.

When it comes to designing active matter systems with the aim of producing an intelligent material, there are two major routes that are being investigated at the moment. The first route is to design the active matter system directly such that its behavior solves a given task. To make this feasible, extensive insight into the dynamics of the specific active system in question is required as the connection between the properties of the active agents and the collective behavior is hard to predict in general. The second route is to use the active matter system as the basis for a (physical) reservoir computing approach, which is an established computational archetype in machine learning. In this approach, the active matter system does not solve the task at hand directly, but rather serves as strongly nonlinear dynamical system that maps low-dimensional input signals to a high-dimensional space of internal states. This high-dimensional state can then be further processed by, e.g., linear classification to achieve complex tasks.

In this book chapter, we will first discuss existing advances and design frameworks for active particle systems solving concrete tasks and discuss the principal challenges of this approach that still remain. Next, we will discuss the reservoir computing approach for active particle systems and review successful implementations of this kind. Following this, we will discuss the potential of optically and acoustically driven particles in the context of reservoir computing. Finally, we will summarize the state of the field and give an outlook to future developments.

\section{Emergent computing with active particles}
Designing active particles to solve a specific task can be seen as a special case of \textit{emergent computing}, i.e., a decentralized computational paradigm of locally interacting agents giving rise to more abstract global behavior that can be interpreted as computation. Most results in this field use specialized cellular automata as the theoretical backbone of designing systems for specific tasks \cite{Crutchfield1995, Olson1995}. While these automata are typically designed to emulate the motion of active agents (see, e.g.,  Refs. \cite{Arroyo2018,Daymude2023}), it is not immediately clear how they can be translated to the continuous particle models typically analyzed in the field of active matter. One can also look at the problem from the other side and start with a typical active particle models such as the Vicsek model \cite{Vicsek1995} and then attempt to tune the model parameters to achieve a formulated goal. However, for more complex models, such as those with spatiotemporal variations in agent activity, the dimensionality of the parameter space becomes large enough that advanced techniques for multidimensional optimization must be employed as brute-force approaches fail. In Ref. \cite{Falk2021}, Falk et al.\ demonstrate that machine learning techniques such as reinforcement learning can be used effectively to explore such a high-dimensional space of parameters and gain physical insights.

With regard to the four functional components of intelligent matter, the sensing and actuation components come naturally with the paradigm of active particles as they utilize (and therefore sense) energy fluxes in their environment and disturb the environment through their actuation, e.g., by inducing flows in the medium or by pushing objects in the environment. Additionally, there are typically strong interactions between active particles (e.g., if two particles collide with opposing directions) which enables signal propagation within the active particle system, e.g., in the form of acoustic waves \cite{teVrugt2021_jerky}. However, the implementation of a memory component in active particle systems is more difficult.  Embedding per-particle artificial memory within microparticles transcends current (mass-)manufacturing technology, so other methods of realizing memory have to be found. One solution is to exploit the inertia of each particle to store information about its past states. It was found that including inertia into consideration of active matter strongly influences the possible collective phases \cite{Caprini2022,Nagai2015}, allows for properties not seen in overdamped active matter such as different modes of sound propagation \cite{teVrugt2021_jerky} or a classical analogue of the tunnel effect \cite{teVrugt2023_schroedinger}, and even helps explain the motion patterns observed in motile microorganisms \cite{MayerMartins2022}. While using inertial memory is convenient, it is as of yet unclear if it is sufficient for the requirements of constructing intelligent matter systems. An alternative approach for \textit{collective} memory storage is presented by Couzin et al.\ in Ref. \cite{Couzin2002}. Here, the authors describe how information can be stored on the group-level by exploiting hysteresis effects in the transition between collective dynamical states.

To the best of our knowledge, there is as of yet no demonstration of an intelligent system of active particles as current implementations have not yet shown the ability for autonomous learning, i.e., they rely on external training to solving a specific task. However, there are numerous promising results for active particle systems showing adaptivity. In the interest of brevity we only give some examples here and refer to the review of Li et al.\ in Ref. \cite{Li2022} for a more exhaustive overview of colloidal self-assembly. For one, there are efforts to use magnetic nanoparticles as the building blocks of adaptive microswarms. An impressive result for this kind of system is given by Yu et al.\ in Ref. \cite{Yu2018} where they show a microswarm navigating through complex boundary geometries. A general overview of the field of magnetic active matter systems with functional collective behavior can be found in the account of Jin et al.\ in Ref. \cite{Jin2022}. A second example are active particle systems showing phototaxis, i.e., a directed motion towards or away from a light source, a common motif found in biological systems as well. In Ref. \cite{Dai2016}, Dai et al.\ present a tunable microswimmer system propelled by light-induced self-electrophoresis that shows collective phototaxis similar to green algae. Mou et al.\ demonstrate in Ref. \cite{Mou2019} how a phototactic flock of TiO$_2$ micromotors can perform collective tasks such as transport of cargo significantly larger than the micromotors. In a yet more complex setup, Liang et al.\ show in Ref. \cite{Liang2020} that phototactic microswarms of unequal individuals can self-organize into hierarchical structures which allows for implementing multiresponsiveness and thus further increases the design possibilities of this class of active particles. It should be noted that phototaxis is not the only taxis behavior possible as shown in the review by Ji et al.\ in Ref. \cite{Ji2023}.

To conclude this section, we also wish to point out the field of \textit{swarm robotics} as a relevant adjacent field of research. While robots are rarely called active particles, the challenge of solving complex tasks collectively with a very limited budget for the complexity of each swarm member is remarkably similar to the field of active particles as we have discussed above. An introduction to the field of swarm robotics can be found in Ref. \cite{Khaldi2015}. In Ref. \cite{Savoie2019}, Savoie et al.\ explore how a swarm of ``smarticles'' (smart active particles in the form of robots following simple rules) can perform higher-level functions such as phototaxis and thus effectively act as a larger robot without a central control unit. Another recent example can be found in Ref. \cite{Saintyves2024} where Saintyves et al.\ present ``granulobots'' as a bridge between soft robotics and active granular materials. These robots take the form of a self-actuating roller that can magnetically connect to its neighbors. Using only local interactions with their neighbors, they exhibit a wide variety of possible collective phases from rigid self-assembly to fluid-like states.

\section{Physical reservoir computing with active matter}
Reservoir computing was developed as a framework for simplifying the training process of recurrent neural networks (see the chapter by te Vrugt for a general introduction to this topic). The key idea is to split the network into a \textit{reservoir} part that nonlinearly maps its input to a large space of internal states and a much simpler \textit{readout layer} (often a simple linear classification) that computes the overall network output by observing the output of the reservoir. Training is then only performed on the readout layer in the hope that classification of inputs is easier in the high-dimensional state space of the reservoir. In \textit{physical} reservoir computing, the reservoir part of the network is  realized as a dynamical system\footnote{The term ``physical reservoir computing'' is somewhat misleading since in principle any time-dependent system is a candidate for physical reservoir computing, even if the dynamical system does not describe an actual physical system. However, since unphysical dynamics require some form of costly emulation (typically on conventional computing hardware), the potential of physical reservoir computing for more efficient computation is only realized when using physical dynamics.}. There is a wide variety of examples of dynamical systems that can be used as reservoirs such as waves on a water surface \cite{Fernando2003}, mechanical networks of anharmonic oscillators \cite{Coulombe2017}, or DNA-based reaction networks \cite{Goudarzi2013}. However, not just any dynamical system is also an  \textit{effective} reservoir. In Ref. \cite{Konkoli2018}, Konkoli et al.\ illuminate the challenge of designing a good reservoir and give an overview of successful implementations. In particular, they describe how a reservoir must ensure that all input states that are to be classified differently by the network must produce distinct internal states in the reservoir (\textit{separation property}). Additionally, a physical reservoir must fulfill the \textit{echo state property}, i.e., it must ensure that earlier inputs eventually fade from the memory of the system. Other in-depth reviews of physical reservoir computing are given by Tanaka et al.\ in Ref. \cite{Tanaka2019} with an even more detailed overview of current realizations and by Nakajima in Ref. \cite{Nakajima2020} with a focus on using soft robotics as a basis for physical reservoir computing. Unfortunately, the research area of physical reservoir computing using active matter systems as the dynamical system for their reservoir is still in its infancy. In the following, we present three promising results that highlight the potential of this direction of research. It should be noted that while these approaches do in principle contain all the functional components for an intelligent system -- sensing and actuation as part of the active particle paradigm as well as a network and memory as part of the reservoir computing paradigm -- there is as of yet no implementation of an actual intelligent system using this technique since all results so far use external training.

First, in Ref. \cite{Lymburn2021}, Lymburn et al.\ present an active matter reservoir computing approach using a modified Reynolds boids model with predator avoidance. The motion of the predator serves as the information input to the system and is chosen to follow the dynamics of the Lorenz system \cite{Lorenz1963} in the chaotic regime. The reservoir computer is then trained to predict future positions of the predator. It was found that using the particle coordinates directly as reservoir output variables does not yield good predictions as such an output definition is inconsistent with the fact that in a system of identical particles any permutation of particles does not change the physical state of the system. To solve this, the authors employed an additional observational layer that observes in random, but fixed locations the state of nearby particles using a Gaussian kernel. With this correction, the authors were able to significantly improve the accuracy of the prediction. It was also observed that using the repulsion force as a control parameter, a phase transition between ordered flock states and disordered states can be triggered and that the reservoir performs optimally when close to this phase transition.

\begin{figure}
    \centering
    \includegraphics[width=\textwidth]{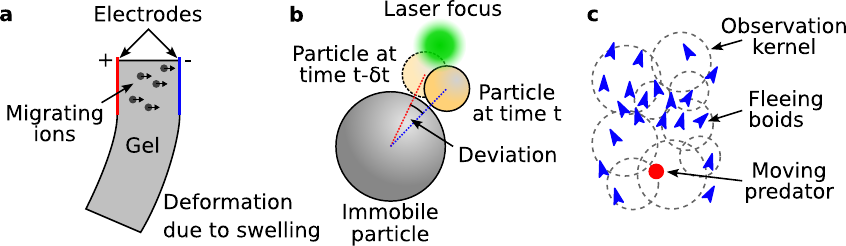}
    \caption{Schematic setups for active particle reservoir computing platforms. \textbf{a} Setup used by Strong et al. in Ref. \cite{Strong2022}. Here, a gel block made from an electroactive polymer is deformed due to the osmotic swelling caused by ionic migration induced by applying an electric field. The setup is submerged in a liquid tank not shown here. Data is fed into the system by modulating the applied voltage and the reservoir output is derived from the degree of deformation of the gel block. \textbf{b} Setup used by Wang et al. in Ref. \cite{Wang2024}.  The particle shown in yellow can move freely in the $xy$-plane and is driven by laser-induced thermophoresis towards an immobilized particle of larger size. The laser focus follows the particle trajectory with a delay $\delta t$ thus causing the active particle to rotate about the immobile particle. The input information is encoded by modulating the laser focus position and information is extracted from the system by observing the time evolution of the deviation angle between the particle positions at times $t-\delta t$ and $t$. \textbf{c} Setup used by Lymburn et al. in Ref. \cite{Lymburn2021}. Here, the system consists of a two-dimensional swarm of boids that are evading a single predator moving through the system. Information is fed into the system in the form of the predator's trajectory and extracted by observing the boid densities through a set of randomly placed Gaussian kernels. }
    \label{fig:fig2}
\end{figure}

Another example for active matter reservoir computing is given by Wang et al.\ in Ref. \cite{Wang2024}. Here, the dynamics of a single active particle is used as the foundation for constructing a reservoir computer. In Fig. \ref{fig:fig2}, the particle setup is shown schematically. The motion of the active particle is governed by a nonlinear delay dynamics. Importantly, this dynamics fulfills the echo state property, i.e., the system has fading memory. To obtain a sufficiently large reservoir for complex tasks, multiple physical setups are run in parallel and time-multiplexing is used to construct multiple virtual nodes for each physical system. With a reservoir constructed from these virtual nodes it was possible to predict the chaotic Mackey-Glass series with good accuracy. The authors note that the noise naturally present in the dynamics of the particle has severely detrimental effects to the function of the reservoir computer. This was mitigated by including past reservoir states in the output layer which significantly improved the reservoir computing performance.

The final example is given by Strong et al.\ in Ref. \cite{Strong2022}. Here, the active agents are hydrogen ions in an electroactive polymer gel (EAP) driven by an external electric field. The resulting ion migration causes osmotic flows which in turn cause localized swelling and deformation of the material. Notably, this deformation is subject to hysteresis effects such that each consecutive deformation decreases the magnitude of future deformations. The dynamics of EAP deformations thus exhibits inherent memory effects. Additionally, the deformations possess intrinsic variability and as such can only be modelled probabilistically. To use EAPs as computing devices, the authors developed encoding schemes for input data as discrete stimulation voltages and for gel deformations as output data. With these they were able to model the EAP behavior in terms of probablistic Moore automata, a probabilistic extension of finite state automata. This was then further extended to reservoir computing by recording the sequence of deformation states of the gel as the input voltages are applied and using this sequence as reservoir output. It is noteworthy that the role of the active particles in this approach is only to provide a nonlinear response of the electroactive polymer, so their dynamics is only implicitly relevant for the reservoir construction, in contrast to the other two examples we gave in this section.

\section{Propulsion mechanisms for active particle reservoir computing}
\begin{figure}
    \centering
    \includegraphics[width=\textwidth]{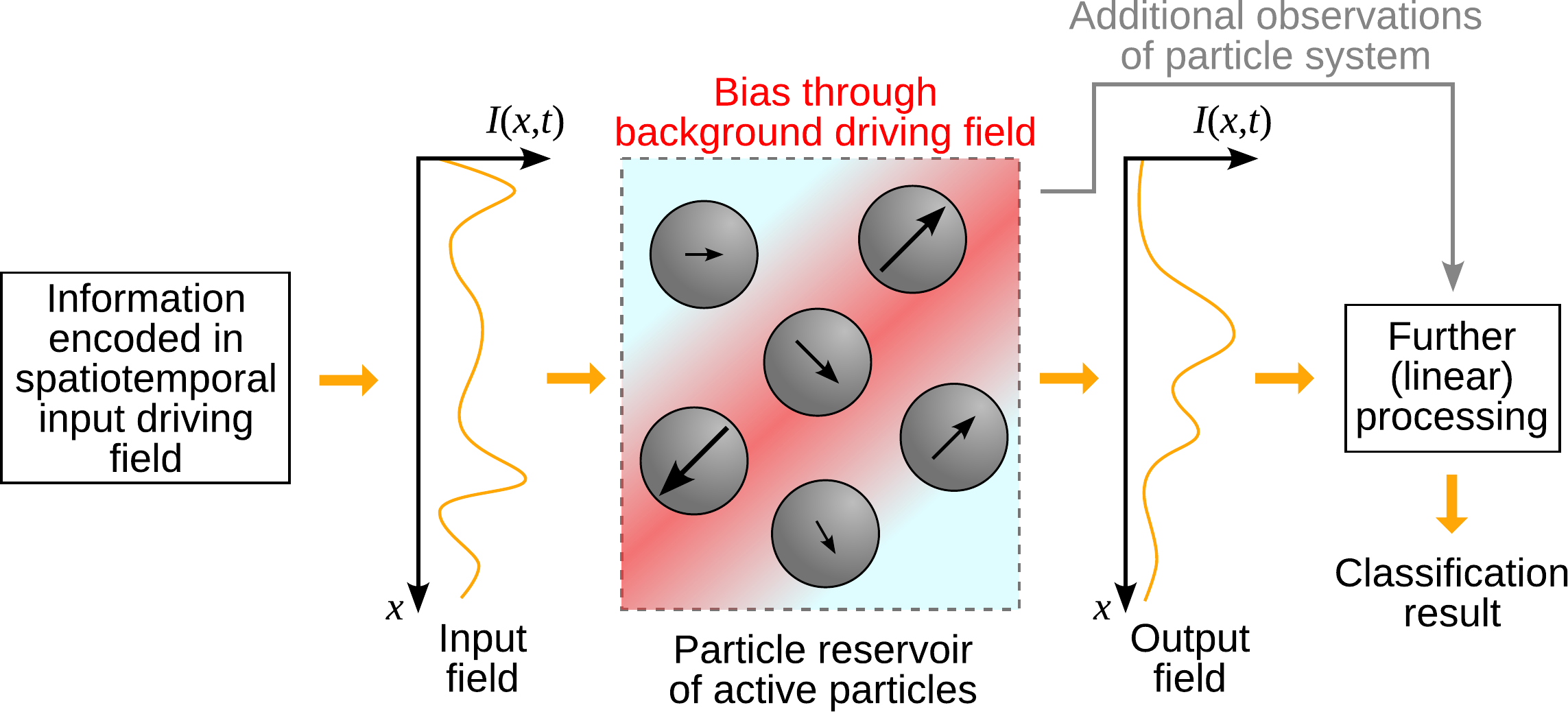}
    \caption{A possible setup for implementing reservoir computing with bulk active particle systems driven by spatiotemporally modulated driving fields. In this example, we show information encoding in the intensity $I(x,t)$ of the input driving field, but one can also imagine information encoding in other quantities affecting the particle propulsion, e.g., polarization for optically driven particles. To achieve greater freedom in the design of the information encoding scheme for the input light field, the particles are also subject to a bias light field with fixed spatiotemporal structure. The readout layer of this reservoir computing scheme is fed with a measurement of the driving fields after they have passed through the system and optionally with other observations of the particle system (e.g., density or velocity averages of particles in specific regions of the reservoir). This scheme can be used for particle reservoirs in two or three spatial dimensions.}
    \label{fig:fig3}
\end{figure}

As noted in the previous section, the topic of using active particle systems to drive reservoir computing remains largely unexplored as of today. We have seen in the few results pioneering this research field that designing reservoirs with high information bandwidth, i.e., large numbers of nodes, is difficult. Thus, implementations often resort to time-multiplexing to construct virtual nodes at the cost of serializing information processing. Furthermore, we have seen that inputting and extracting information from an active particle reservoir requires dedicated schemes for data encoding to achieve good reservoir performance. Finally, we have discussed that memory design is critical for the design of a good reservoir due to the echo state property. However, active particle systems often introduce memory effects as an unwanted side product of the propulsion mechanism. For example, chemically fueled particles will generally leave concentration gradients in their wake and will eventually deplete the fuel in the system \cite{Liebchen2017,Hokmabad2021} while bulk systems of thermophoretic particles can incidentally introduce memory in the form of the average system temperature \cite{Auschra2021}. In this section, we will give a description of two propulsion schemes for active particles that we expect to mitigate these issues, but that have not been used for the purpose of reservoir computing to the best of our knowledge. In Fig. \ref{fig:fig3}, we sketch a possible setup for using these propulsion mechanisms in a reservoir computing context. The core idea behind this setup is to encode information in the spatiotemporal structure of the input driving field and thus achieving high information bandwidth in the reservoir. So far, there have been no realizations of such a system and as such its efficacy remains speculative.

\begin{figure}
    \centering
    \includegraphics[width=\textwidth]{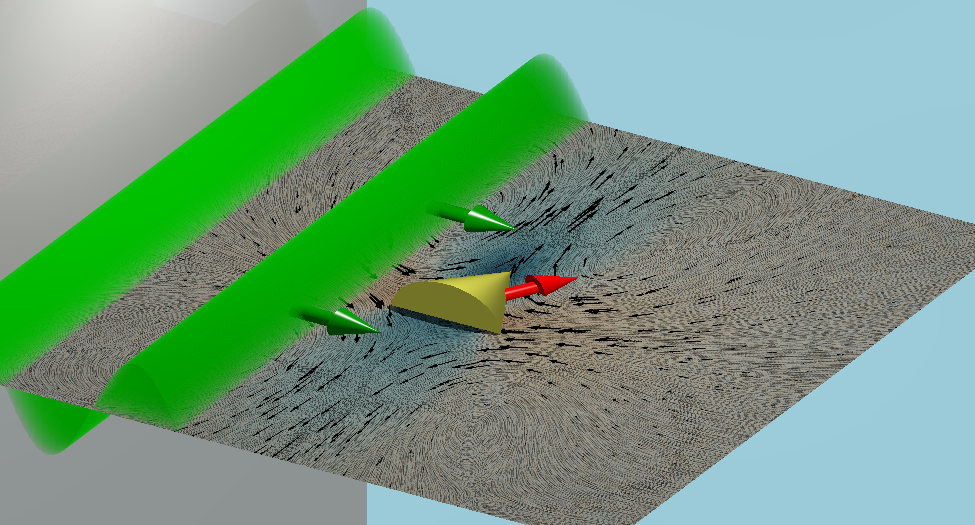}
    \caption{Illustration of a planar ultrasound wave (shown schematically in green) interacting with a microparticle submerged in a fluid. Boundary-driven acoustic streaming creates a flow field around the particle driving it in the direction indicated by the red arrow. The wavelength of the ultrasonic field was shortened significantly for illustrative purposes and would be much larger than the particle if shown to scale. This image was created by Johannes Vo{\ss} and Raphael Wittkowski, published in Ref. \cite{Voss2022} under the  Creative Commons Attribution 3.0 Unported Licence (see \textit{creativecommons.org/licenses/by/3.0/)}, and is reproduced here without changes.}
    \label{fig:fig4}
\end{figure}

\begin{figure}
    \centering
    \includegraphics[width=0.92\textwidth]{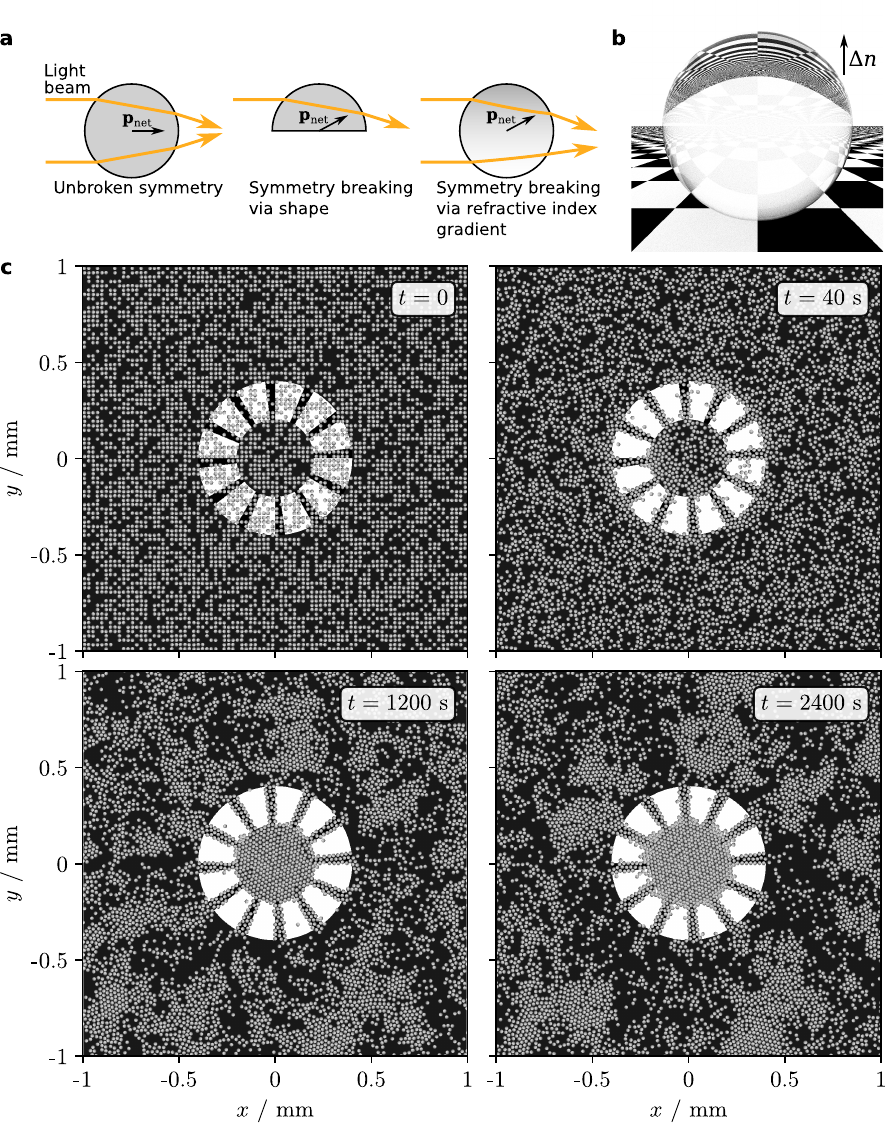}
    \caption{Overview of symmetry-broken refractive microparticles. \textbf{a} Sketch showing the principle behind the refractive propulsion. Momentum conservation dictates that, in general, light refraction transfers momentum $\vec{p}_\textrm{net}$ from the light field to the refracting material. To achieve orientation-dependent propulsion, the particle needs to break symmetry either in shape or in the refractive index profile (here symbolized by a color gradient). \textbf{b} Raytracing rendering of a particle with a symmetry-breaking refractive index gradient (the direction is denoted by $\Delta n$). The center of the viewport is on the same height as the center of the spherical particle. The distorted image of the horizon thus demonstrates the breaking of symmetry. \textbf{c} Dynamic simulation of a system of refractive microparticles in two spatial dimensions illuminated perpendicular to the particle domain. The illumination intensity is shown as the background color with white denoting a high intensity and black denoting a low intensity. It can be observed that the higher propulsion velocity in the bright region displaces particles into the darker regions. With the illumination structure used here, this leads to the formation of walls made of slowly moving particles that funnel incoming particles to the center of the illumination pattern and the associated growth of a crystal in the center of the domain. Notably, this crystal also grows into the areas of high illumination intensity. This behavior is an indication of \textit{adaptivity} as the overall particle distribution is no longer the simple superposition of the reaction of each particle to the local illumination intensity, but rather we observe a nonlocal dependency of the average particle density on the illumination pattern. This shows that the particle interactions form a network capable of exchanging information between different parts of the system.}
    \label{fig:fig5}
\end{figure}

The first propulsion mechanism useful in this context is based on acoustically driven particles \cite{Li2022_acoustic,Ren2022,McNeill2023}. In this type of active matter, ultrasonic pressure waves interact with microparticles which induces flows in the medium around the particle driving the particles forward. An illustration of this principle is provided in Fig. \ref{fig:fig4}. Notably, the acoustic propulsion does not rely on creating secondary gradients (e.g., in concentrations or temperature) in the medium and thus does not introduce secondary memory effects. An ultrasonic field can be spatially modulated and recorded with high resolution as has recently been demonstrated by Ma et al.\ \cite{Ma2020}. The design of the propulsion offers many degrees of freedom since the acoustic propulsion is determined by the shape of the particles \cite{Voss2020} and their material which can be tuned by microfabrication techniques. Possible designs include nanorods \cite{Li2022_acoustic,Ren2022,McNeill2023}, particles including microbubbles \cite{Bertin2017}, and screw-shaped microrobots \cite{Deng2023}. With these designs, a wide range of propulsion properties can be realized (e.g., reversal of the propulsion direction by tuning the aspect ratio \cite{Voss2022}). In combination, these properties make acoustically propelled particle systems a promising candidate for active particle reservoir computing.

The second propulsion mechanism we present uses the momentum transfer that occurs on light refraction to generate directed motion in microparticles with a symmetry-broken refractive index profile (SBRIP) \cite{Volpe2023,Denz2019,Denz2021} as shown in Fig. \ref{fig:fig5} \textbf{a} and \textbf{b}. These particles can be produced experimentally by microfabrication techniques such as two-photon polymerization \cite{Rueschenbaum2021}. Such a system can then be used as a building block for all-optical reservoir computing. This is particularly promising as photonic systems have been one of the most successful approaches to physical reservoir computing so far \cite{Tanaka2019,VanDerSande2017}. One of the principal difficulties in these types of systems is the requirement for nonlinearity in the reservoir. By coupling the dynamics of an active particle system to the light field, this requirement for nonlinearity can be fulfilled without compromises like the time-multiplexing used in delay-based photonic reservoirs which necessitates serialization of the input and output data. When it comes to driving active matter systems optically, a refraction-based approach has multiple advantages. Since the light does not interact with the medium directly, side effects (e.g., concentration gradients) from the propulsion are avoided. Compared to absorptive or reflective propulsion methods, refraction allows for a much larger attenuation length of the light thus permitting active bulk systems. When using the refractive index profile of the particles to achieve the necessary breaking of symmetry for propulsion, one can effectively decouple shape-dependent hydrodynamic and steric particle interaction properties from the propulsion mechanism, thus allowing for a much greater freedom in particle design. In Fig. \ref{fig:fig5} \textbf{c}, it is demonstrated that the extensive possibilities to modulate light fields in space and time can be used to tailor system behavior. Even the simple case of a two-dimensional system of active particles with position-dependent propulsion speed can show adaptive system behavior by creating feedback between steric particle interactions and the particle propulsion: The propulsion dynamics leads to particle accumulation in the areas of lower propulsion speed. When high particle densities are reached in these areas, new particles are stopped from entering them. This causes the particles to drift along the edge of the dark region instead and thus effectively alters the propulsion velocity in this region. In bulk systems, the possible complexity grows further as then the light field itself becomes coupled to the particle dynamics: The light field induces changes in the particle density which in return changes the light field as it passes through the system. Furthermore, the reservoir can be used in an optical pipeline with other photonic elements such as mirrors or delay lines to introduce additional feedback. Overall, these properties make SBRIP particles a promising platform for the development of physical reservoir computing.

It should be noted that while we did outline the propulsion mechanisms in this section for the purpose of reservoir computing, they can, of course, also be used for the task of emergent computing. To the best of our knowledge, this direction has not been explored either so far. Therefore, we cannot make a statement on the efficacy of this approach.

\section{Conclusions and outlook}
In summary, we have shown how active matter systems are a promising platform for constructing intelligent matter due to their inherently nonequilibrium nature that allows complex collective behavior patterns. While truly intelligent active matter systems have not been realized yet, there are results showing advanced adaptive behavior in these kinds of systems. In particular, there is a growing field of emergent computation with active particles where the individual particles are following simple rules and yet produce a swarm behavior capable of solving useful tasks. However, a large challenge that remains in this domain is the implementation of memory which is necessary for the step to intelligent systems.

We have also seen that there are efforts to implement physical reservoir computing using active particle systems as the underlying dynamical system. While this field is very much in its infancy, the first successful results seem to indicate that this approach works in principle. One of the major challenges still open in this field is to integrate the learning procedure into the system itself (e.g., by integrating a reinforcement learning scheme into the physical system), which is considered a requirement for intelligent matter systems.  The success of the reservoir computing approach begs the question whether active particle systems can also be used to emulate other machine learning archetypes, e.g., to emulate neural networks by networks of active particle suspensions acting as neurons. Indeed, there remains much work to be done to discover the full potential of active matter systems for the construction of artificial intelligence.

\section{Acknowledgements}

This work is funded by the Deutsche Forschungsgemeinschaft (DFG, German Research Foundation) –- Project-ID 433682494 -- SFB 1459.

%
%
\bibliographystyle{spphys}
\bibliography{literature}

\end{document}